\documentclass{sigchi-ext}
\usepackage[T1]{fontenc}
\usepackage{textcomp}
\usepackage[scaled=.92]{helvet} 
\usepackage{graphicx} 
\usepackage{balance}  
\usepackage{booktabs} 
\usepackage{ccicons}  
\usepackage{ragged2e} 
 \usepackage{url}

\def\plaintitle{SIGCHI Extended Abstracts Sample File: Note Initial
  Caps} 
\def\emptyauthor{}
\def\plainkeywords{virtual reality; keyboard; mouse; immersive analytics; head-mounted displays}

\title{Back to the Future: Revisiting Mouse and Keyboard Interaction for HMD-based Immersive Analytics}

\numberofauthors{4}
\author{%
  \alignauthor{%
    \textbf{Jens Grubert}\\
    \affaddr{Coburg University of Applied Sciences and Arts} \\
    \email{jens.grubert@hs-coburg.de} } \alignauthor{%
    \textbf{Eyal Ofek}\\
    \affaddr{Microsoft Research}\\
    \email{eyalofek@microsoft.com} } \vfil \alignauthor{%
    \textbf{Michel Pahud}\\
    \affaddr{Microsoft Research}\\
    \email{mpahud@microsoft.com} }\alignauthor{%
    \textbf{Per Ola Kristensson}\\
    \affaddr{University of Cambridge}\\
    \email{pok21@cam.ac.uk} } \vfil 
    }

\definecolor{linkColor}{RGB}{6,125,233}
\hypersetup{%
  pdftitle={\plaintitle},
  pdfauthor={\emptyauthor},
  pdfkeywords={\plainkeywords},
  bookmarksnumbered,
  pdfstartview={FitH},
  colorlinks,
  citecolor=black,
  filecolor=black,
  linkcolor=black,
  urlcolor=linkColor,
  breaklinks=true,
}

\begin{document}

\CopyrightYear{2020}
\setcopyright{rightsretained}
\conferenceinfo{CHI'20,}{April  25--30, 2020, Honolulu, HI, USA}
\isbn{978-1-4503-6819-3/20/04}
\doi{https://doi.org/10.1145/3334480.XXXXXXX}
\copyrightinfo{\acmcopyright}

\maketitle

\RaggedRight{} 

\begin{abstract}
With the rise of natural user interfaces, immersive analytics applications often focus on novel forms of interaction modalities such as mid-air gestures, gaze or tangible interaction utilizing input devices such as depth-sensors, touch screens and eye-trackers. At the same time, traditional input devices such as the physical keyboard and mouse are used to a lesser extent. We argue, that for certain work scenarios, such as conducting analytic tasks at stationary desktop settings, it can be valuable to combine the benefits of novel and established input devices as well as input modalities to create productive immersive analytics environments.
\end{abstract}






\section{Introduction}


The area of Immersive Analytics tries to remove barriers between data, people who analyze this data and the tools they use to do so \cite{marriott2018immersive}. Researchers combine knowledge from fields such as data visualization, human-computer interaction and mixed reality to create and study new tools and approaches to engage with data. The rise of natural user interfaces as well as the introduction of affordable immersive head-mounted displays (HMDs) \cite{cliquet2017towards} led to a wide variety of interaction techniques for data and view specification and manipulation \cite{bowman20043d, heer2012interactive} including touch, spatial gestures, tangible and gaze interaction and a number of archetypal setups such as large screen collaborative spaces (with or without personal displays such as tablets) or immersive setups (projection or head-mounted display-based) (for an overview we refer to Büschel et al. \cite{buschel2018interaction}).

Specifically, HMD-based systems make heavy use of spatial gestures using bare hands or controllers but are typically designed to support free-space interaction, assuming no interfering objects or humans nearby. While this allows for expressive, and potentially co-located interaction, free space  interaction comes at the cost of increased fatigue \cite{hincapi2014consumed} or inaccurate input (e.g., when using hand or gaze-based ray casting techniques \cite{brown2014performance, qian2017eyes}). While a number of techniques have been proposed to facilitate object selection in presence of clutter (e.g.,  \cite{sidenmark2020outline}), to increase spatial pointing accuracy \cite{argelaguet2013survey, kyto2018pinpointing} or to mitigate fatigue of spatial gestures \cite{hansberger2017dispelling} they still do not eliminate those challenges.

We argue, that the combination of desktop-based input devices such as the physical keyboard and mouse with immersive head-mounted displays can benefit single users in immersive analytics tasks, similar to office-based knowledge work \cite{grubert2018office, citi2016} or the use of hybrid 2D/3D interaction in medicine \cite{mandalika2018hybrid}.


\begin{marginfigure}[-40pc]
  \begin{minipage}{\marginparwidth}
    \centering
     \includegraphics[width=\linewidth]{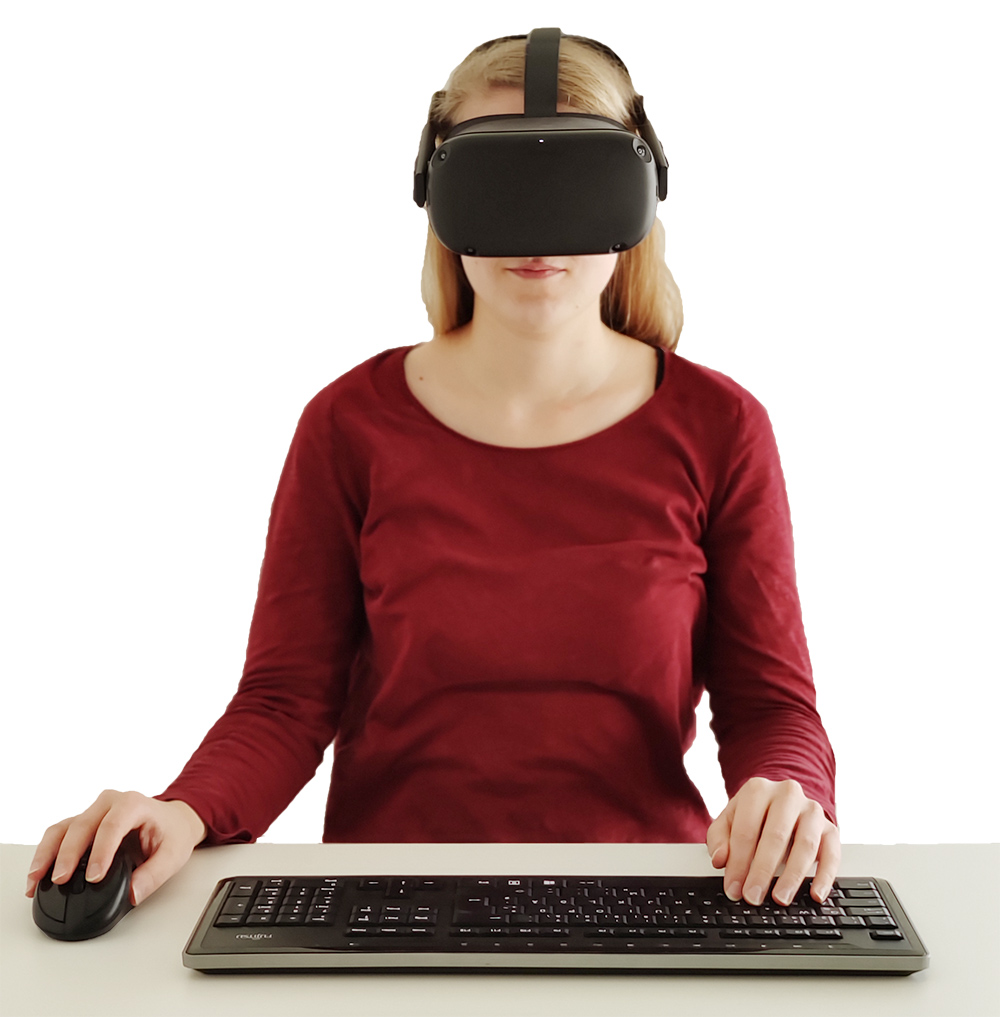}
   \caption{Interaction with head-mounted display, keyboard and mouse.}
   \label{fig:overview}
  \end{minipage}
\end{marginfigure}

\begin{marginfigure}[-12pc]
\begin{minipage}{\marginparwidth}
  \centering
  \includegraphics[width=\linewidth]{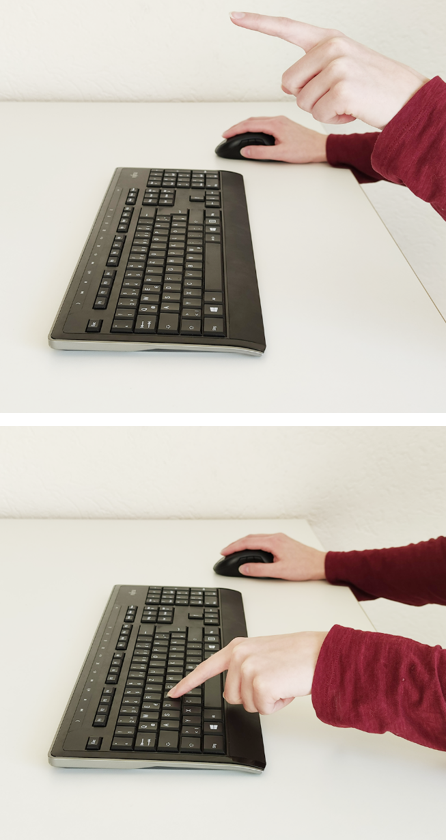}
  \caption{Top: 3D pointing with one hand, selection confirmation via mouse press. Bottom: Transition from mid-air pointing to key press.}
  \label{fig:3d22d}
    \end{minipage}
\end{marginfigure}

\section{Keyboard and Mouse for HMD-based Immersive Analytics}

The physical keyboard and mouse are optimized for symbolic and precise 2D input and have a long tradition in being used as standard input devices in desktop environments. While not free from challenges, they have been optimized to support long hours of work \cite{brand2008office, woo2016ergonomics}. 

The keyboard was designed for rapid entrance of symbolic information, and although it may not be the best mechanism developed for the task, its familiarity that enabled good performance by users without considerable learning efforts kept it almost unchanged for many years. However, when interacting with spatial data, they are perceived as falling short of providing efficient input capabilities \cite{besanccon2017mouse}, even though they are successfully used in many 3D environments (such as CAD or gaming \cite{stuerzlinger2011value}), can be modified to to allow 3D interaction \cite{ware1997selection, perelman2015roly} or can outperform 3D input devices in specific tasks such as 3D object placement \cite{berard2009did, sun2018comparing}. 

With the advent of self-contained immersive head-mounted displays, which allow for spatial tracking of the environment and the users hand, as well as eye-tracking, there is a potential to efficiently utilize keyboard and mouse interaction in single user, desktop-based environments (see Figure \ref{fig:overview}) for immersive analytics tasks. For example, Wang et al. \cite{wang2020towards} explored the use of an Augmented Reality extension to a desktop-based analytics environment. Specifically, they added a stereoscopic data view using a HoloLens to a traditional 2D desktop environment and interacted with keyboard and mouse across both the HoloLens and the desktop. Furthermore, the ability of immersive near eye displays to modify the visual representations of keyboard and mouse enhance their flexibility allows for application-specific adaptations \cite{schneider2019reconviguration}. 

Along this research trajectory, we see the following aspects applicable to immersive analytics using virtual reality or video see-through-based augmented reality. \\

\subsection{Complementary and Multi-modal Input} So far, problems in switching between spatial interaction (e.g., using controllers) and keyboard and mouse interaction have limited the applicability of  desktop-based input devices for immersive analytics. Even in stationary, desktop-based scenarios it might be challenging to switch from motion-tracked controllers to keyboard and mouse devices. However, given the possibility to spatially track the users hands and the keyboard and mouse through model-based tracking \cite{lepetit2005monocular, marchand2015pose} applicable to today's HMDs with  camera-based inside-out tracking, we see the potential to seamlessly switch between mid-air interaction and mouse or keyboard input, see Figure \ref{fig:3d22d}. This could open up efficient switching between tasks (e.g., selecting 3D around the user through spatial gestures and changing data properties through symbolic input on the keyboard) or subsequent fine-grained selection on a 2D subspace of the data using the mouse. Further, the input devices can be combined for multi-modal interaction. For example, one hand could be used for (uncertain) data selection again, while the other hand could be used for certain action confirmation, e.g., through key press on the physical keyboard, or alternatively for moving the data views around the user - instead of having the user navigate through the virtual scene. 

\begin{marginfigure}[-29pc]
\begin{minipage}{\marginparwidth}
  \centering
  \includegraphics[width=\linewidth]{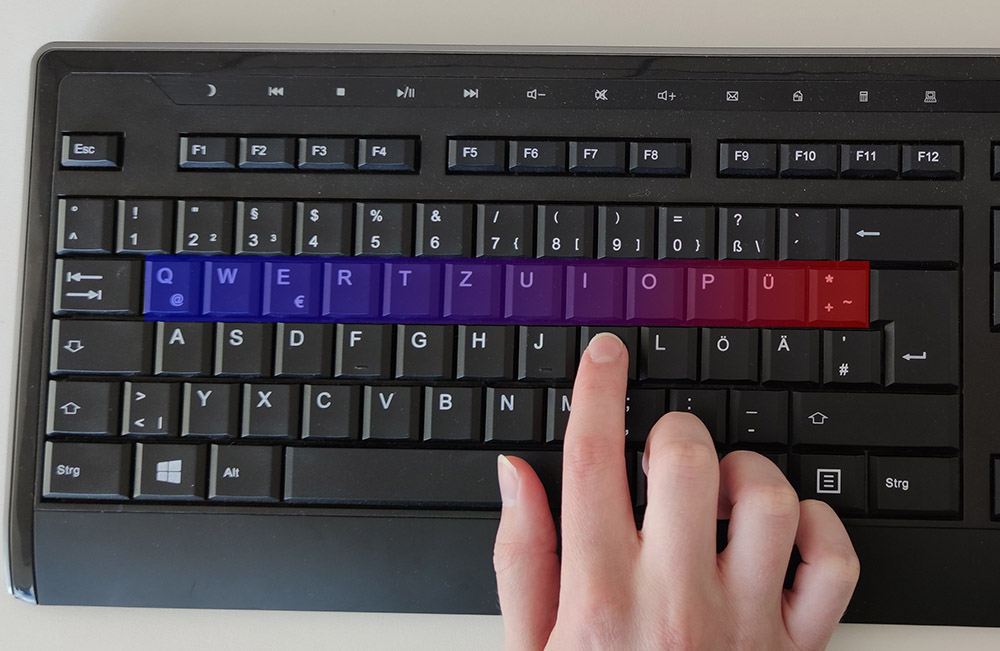}
  \caption{Color scale mapped to keyboard keys. Color selection could be interpolated by pressing two buttons at once.}
  \label{fig:colorscale}
    \end{minipage}
\end{marginfigure}




\subsection{Augmenting peripherals} Virtual data entities can also be augmented on or around the keyboard and mouse to allow for direct interaction with those virtual data items \cite{schneider2019reconviguration}. For example, in a node-link diagram, individual nodes could be associated to individual keys to allow quick selection of individual nodes (i.e. one key is mapped to one data entity), to multiple keys e.g., when only few nodes are present, or a single key could represent multiple nodes (e.g. in a dense node-link diagram with many nodes). Similarly, user interface elements for manipulating object properties, such as sliders could be mapped to multiple keys on the keyboard, to the mouse-wheel or to the area around the mouse. Also, different areas on a physical mouse with touch sensitive surfaces could have different semantics. Again, the advantage of mapping these graphical elements to the physical input devices lies in the increased certainty of the input (e.g., key press, moving the mouse over a physical surface) in contrast to uncertain mid-air or gaze-based input. In addition, a spatially tracked mouse could be utilized to enable constrained 3D object manipulations such as rotations or scaling.


\section{Conclusion and Future Work}
Through this position paper, we aim at increasing the awareness about the potential that traditional desktop-based input devices such as the physical keyboard and mouse can bring into immersive analytics tasks. It lies in the combination of certain but (in terms of degrees of freedom) spatially limited input of those devices with expressive but uncertain and fatiguing spatial input, as well as the ability to virtually augment keyboard and mouse for enhanced interaction in immersive analytics tasks. In future work, we aim at investigating specific immersive analytics tasks and at studying the opportunities of multi-modal interaction between spatial and keyboard and mouse-based interaction in more detail. Finally, we will also explore the  opportunities of integrating stationary touch-screens (e.g. integrated in laptops) for immersive analytics tasks.

\balance{} 

\bibliographystyle{SIGCHI-Reference-Format}
\bibliography{sample}

\end{document}